\documentclass[twocolumn]{aastex631}

\usepackage{url}

\usepackage{amsmath}
\usepackage{float}
\usepackage{rotating}
\usepackage{gensymb}
\usepackage{hyperref}
\graphicspath{{./}{Figures/}}

\shorttitle{Origins of Hot Jupiters from the Stellar Obliquity Distribution}
\shortauthors{Rice et al.}

\begin{document}
\title{Origins of Hot Jupiters from the Stellar Obliquity Distribution}

\author[0000-0002-7670-670X]{Malena Rice}
\altaffiliation{NSF Graduate Research Fellow}
\affiliation{Department of Astronomy, Yale University, New Haven, CT 06511, USA}

\author[0000-0002-7846-6981]{Songhu Wang}
\affiliation{Department of Astronomy, Indiana University, Bloomington, IN 47405, USA}

\author[0000-0002-3253-2621]{Gregory Laughlin}
\affiliation{Department of Astronomy, Yale University, New Haven, CT 06511, USA}

\correspondingauthor{Malena Rice}
\email{malena.rice@yale.edu}

\begin{abstract}
The obliquity of a star, or the angle between its spin axis and the average orbit normal of its companion planets, provides a unique constraint on that system's evolutionary history. Unlike the Solar System, where the Sun's equator is nearly aligned with its companion planets, many hot Jupiter systems have been discovered with large spin-orbit misalignments, hosting planets on polar or retrograde orbits. We demonstrate that, in contrast to stars harboring hot Jupiters on circular orbits, those with eccentric companions follow no population-wide obliquity trend with stellar temperature. This finding can be naturally explained through a combination of high-eccentricity migration and tidal damping. Furthermore, we show that the joint obliquity and eccentricity distributions observed today are consistent with the outcomes of high-eccentricity migration, with no strict requirement to invoke the other hot Jupiter formation mechanisms of disk migration or in-situ formation. At a population-wide level, high-eccentricity migration can consistently shape the dynamical evolution of hot Jupiter systems.
\end{abstract}


\section{Introduction} 
\label{section:intro}

The sky-projected obliquities, $\lambda$, of over 140 exoplanet-hosting stars have been determined to date by observing the Rossiter-McLaughlin effect \citep{rossiter1924detection, mclaughlin1924some} in which a transiting planet blocks out different components of a rotating star’s light as it passes across the stellar profile. These measurements have revealed a diversity of projected angles between the stellar spin axis and the orbit normal vectors of neighboring planets, with systems spanning the full range of possible configurations from prograde to polar and retrograde orbits.
 
Because Rossiter-McLaughlin observations require a transiting geometry and at least 10-12 high-resolution in-transit spectra, they are limited to only a subset of the known population of exoplanets and have typically been made for hot Jupiters -- giant planets on tight orbits. Several channels have been proposed for hot Jupiter formation (see \citet{dawson2018origins} for a comprehensive overview), including (1) high-eccentricity migration, in which planets born on wide orbits reach extremely high eccentricities before tidally circularizing to their current orbits \citep[e.g.][]{wu2003planet, fabrycky2007shrinking, wu2007hot, nagasawa2008formation, beauge2012multiple}; (2) disk migration, in which planets born on wide orbits migrate inwards within the disk plane \citep[e.g.][]{goldreich1980disk, lin1986tidal, lin1996orbital}; and (3) in-situ formation, in which planets form on similar orbits to those on which they currently lie \citep[e.g.][]{batygin2016situ, boley2016situ}. The stellar obliquity distribution may provide compelling evidence to distinguish between hot Jupiter formation mechanisms: hot Jupiters formed through high-eccentricity migration, should commonly attain both high eccentricities and large misalignments early in their evolution. On the other hand, hot Jupiters formed in-situ or through disk migration should typically be aligned in the absence of disk- or star-tilting perturbers, with no requirement to reach high eccentricities or inclinations at any point in their evolution.

 
The primary observational result from previous studies of exoplanet host star obliquities is that hot stars hosting hot Jupiters span a wider range of obliquities than their cool star counterparts \citep{winn2010hot, schlaufman2010evidence}. The transition point occurs at the Kraft break ($T_{\rm eff} \sim 6100$ K), a rotational discontinuity above which stars rotate much more quickly and lack thick convective envelopes \citep{kraft1967studies}. The observed discontinuity in obliquities at the Kraft break is commonly attributed to differences in the tidal realignment timescale for stars above and below the Kraft break \citep{winn2010hot, albrecht2012obliquities}, and alternative explanations invoking magnetic braking \citep[e.g.][]{dawson2014tidal, spalding2015magnetic}, internal gravity waves \citep{rogers2012internal}, and differences in the external companion rate \citep{wang2021soles} have also been proposed. To date, however, this trend has only been demonstrated for the full population of giant-exoplanet-hosting stars with measured obliquities, without delineating the role of the companions' orbital eccentricity, $e$. 


In this work, we show that the population of exoplanets on eccentric orbits reveals no evidence for this well-established transition at the Kraft break. While we recover this discontinuity for the population of exoplanets on circular orbits, it is not present for exoplanets on eccentric orbits. This discrepancy supports high-eccentricity migration as a key hot Jupiter formation mechanism, where the final obliquities of hot Jupiter systems are shaped by tidal dissipation.

\section{Population-Wide Obliquity Analysis}

We compared two populations: ``eccentric" planets with $e\geq0.1$ -- ranging from $e=0.1$ to $e=0.93$ in our sample -- as well as “circular” ($e = 0$) planets with a reported eccentricity of exactly zero. As in \citet{wang2021aligned}, the cutoff for ``eccentric'' planets was set to $e=0.1$ to remove systems with the most modest eccentricities, which may be more analogous to the ``circular'' population due to the Lucy-Sweeney bias \citep{lucy1971spectroscopic, zakamska2011observational}. 

\begin{figure*}
    \centering
    \includegraphics[width=1.0\linewidth]{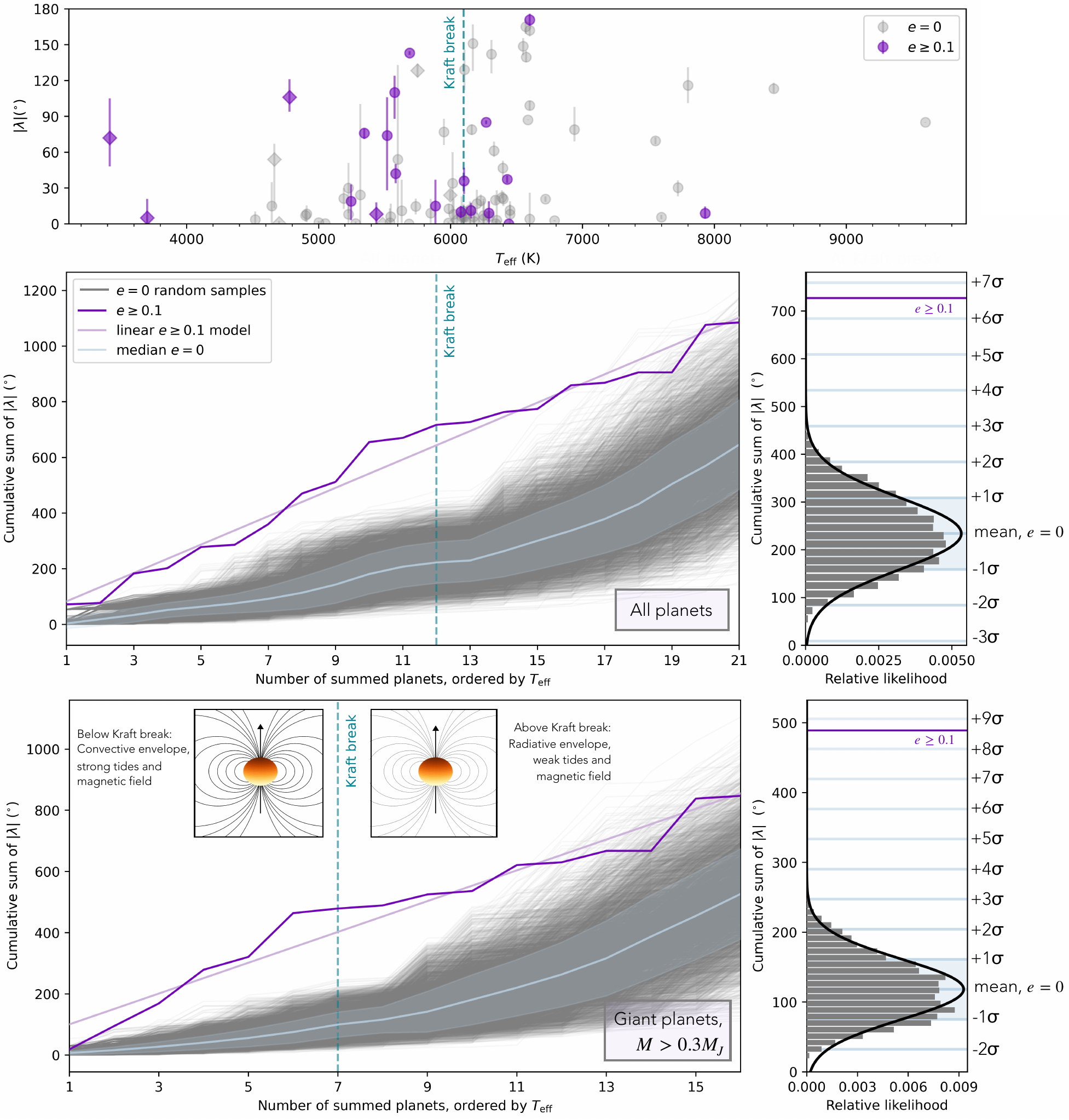}
    \caption{Comparison of the obliquity distributions for stars hosting exoplanets on circular vs. eccentric orbits. Top: Full sample of spin-orbit angles included in this study. Planets with $M < 0.3M_J$ are shown with diamond markers. The data behind this panel is available together with all other planet parameters used in this work, drawn from archival studies. Middle and bottom: Cumulative sums of $|\lambda|$ for eccentric exoplanets, compared with 5000 randomly sampled sets of circular exoplanets (sampled without replacement). Histograms on the right provide vertical cuts through the sums at the Kraft break. In each panel, a linear model fitting the $e \geq 0.1$ cumulative sum is shown in light purple, while the running median of the $e = 0$ population is provided in light blue together with the shaded region within 1$\sigma$ of the median.}
    \label{fig:cumsum_3panel}
\end{figure*}

The samples were drawn from the set of planets included within both the NASA Exoplanet Archive “Confirmed Planets" list and the TEPCat catalogue \citep{southworth2011homogeneous} of sky-projected spin-orbit angles, both downloaded on 10/20/2021. For planets with multiple spin-orbit angle measurements available through TEPCat, only the most recent measurement was included within this analysis, with the exception of systems for which previous observations were much more precise. We included only planets with pericenter distances $q<0.1$ au, which can be tidally circularized on relatively short timescales necessary for high-eccentricity migration. The distribution of obliquities for both populations as a function of stellar temperature is provided in the top panel of Figure\,\ref{fig:cumsum_3panel}.

A set of cumulative sums comparing the eccentric and circular obliquity distributions, where misalignments are accumulated as a function of stellar temperature, reveals that the population-wide change in obliquities at the Kraft break is present for stars hosting planets on circular orbits but does not extend to systems with eccentric orbits (Figure \ref{fig:cumsum_3panel}). The $e\geq0.1$ cumulative sum is shown in purple in Figure \ref{fig:cumsum_3panel}, where the absolute value of the projected spin-orbit angle, $|\lambda|$, is used to uniformly display the deviation of each system from perfect alignment. To compare this result with the $e=0$ population, we divided the circular sample into populations with host star $T_{\rm eff}$ below and above the Kraft break. Random planets were selected from each set to match the number of planets on either side of the Kraft break in the $e \geq 0.1$ population, with 5000 iterations to sample the full parameter space. Then, the $e = 0$ planet samples were ordered by $T_{\rm eff}$ and cumulatively summed. 

We recovered the previously reported trend in obliquity as a function of stellar temperature, confirming that the trend is stronger when excluding lower-mass ($M < 0.3M_J$) planets from the sample. However, this relation holds \textit{only for the population of planets on circular orbits.} At the Kraft break, the eccentric cumulative sum is a $6.5\sigma$ outlier from the circular distribution with all planets included (middle panel of Figure \ref{fig:cumsum_3panel}) and an $8.7\sigma$ outlier from the circular giant planets distribution (bottom panel of Figure \ref{fig:cumsum_3panel}).  

To determine the likelihood that the $e = 0$ trend with stellar temperature is absent for $e \geq 0.1$ planets, we compared two models: a linear model and a running median of the $e = 0$ population, which represents the null result. The fit of each model to the cumulatively summed data was evaluated using the Bayesian Information Criterion \citep[BIC;][]{schwarz1978estimating} and Akaike Information Criterion \citep[AIC;][]{akaike1973maximum} metrics.

The BIC is given by

\begin{equation}
   {\rm BIC} = k\ln N - 2\ln \mathcal{L},
\end{equation}
where $k$ is the total number of parameters estimated by the model, $N$ is the number of planets in the cumulative sum, and $\mathcal{L}$ is the likelihood function. For the likelihood function, we used a reduced $\chi^2$ metric comparing the cumulative sum obtained from the data ($y_{\rm data}$) with the corresponding values for each model ($y_{\rm model}$):

\begin{equation}
    \mathcal{L} = \frac{1}{N}\sum{(y_{\rm data} - y_{\rm model})^2}.
\end{equation}

We adopted the corrected AIC metric (AICc), which includes an adjustment for small sample sizes with $N<40$ \citep{hurvich1989regression}.

\begin{equation}
     {\rm AIC} =  2k - 2N\ln \mathcal{L}
\end{equation}

\begin{equation}
     {\rm AICc} =  {\rm AIC} + \frac{2k(k+1)}{N - k - 1}
\end{equation}

Both the BIC and the AICc strongly favor a linear model over the median $e=0$ cumulative sum (see Figure 1), with $\Delta$BIC$=82$ (52) and $\Delta$AICc$=83$ (53) for the all- (giant-) planets fit. Based on the AICc for each model, the null result, which increases in gradient at the Kraft break, is {$8 \times 10^{-19}$ ($3 \times 10^{-12}$)} times as likely as the linear model to minimize the information loss in the all- (giant-) planet fit. The eccentric and circular populations appear to be distinct.

\begin{figure}
    \centering
    \includegraphics[width=1.0\linewidth]{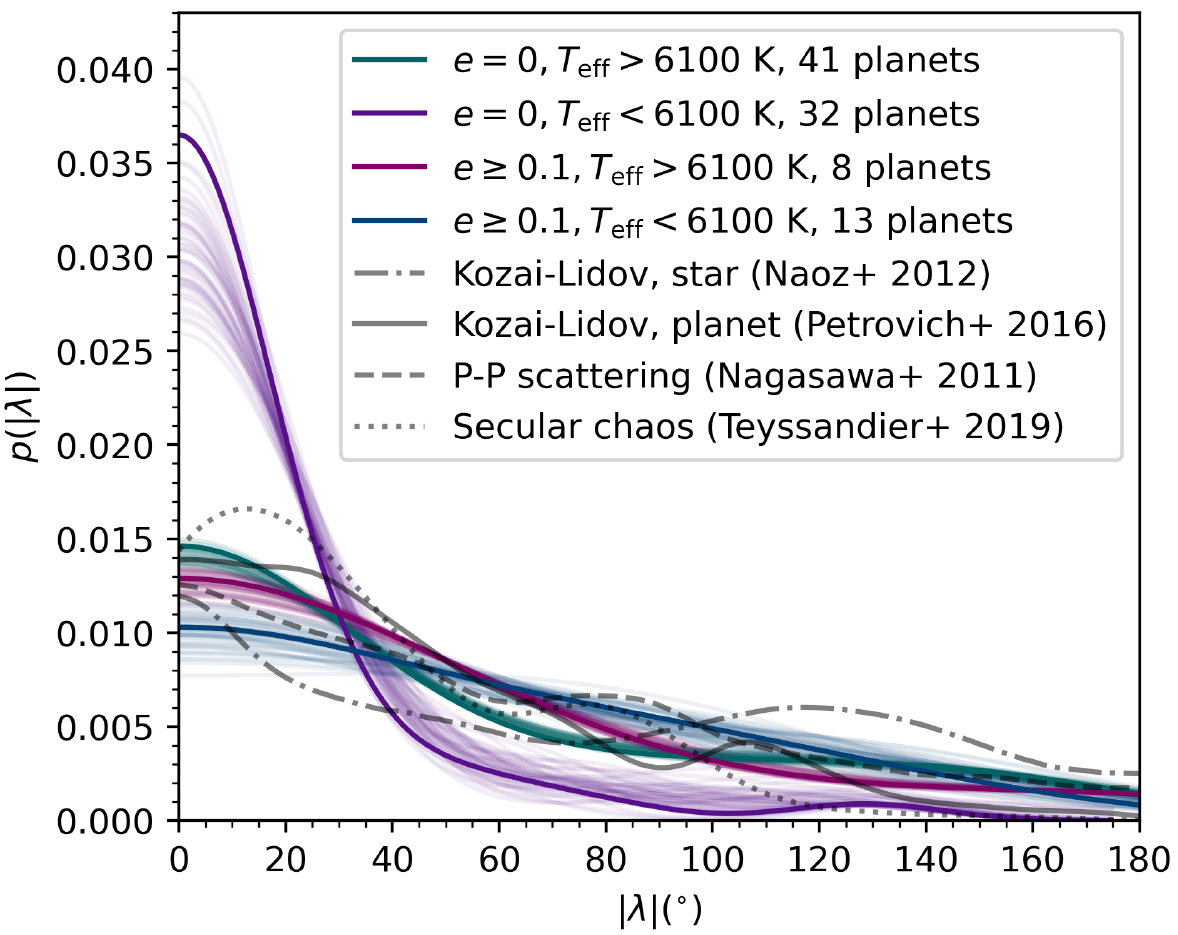}
    \caption{Distribution of $|\lambda|$ values for each population, segmented by eccentricity and host star temperature. The central $|\lambda|$ distribution for each set of planets is shown in bold. To quantify uncertainties, we iteratively drew from a Gaussian distribution around each measured $|\lambda|$ value and recalculated the probability distribution based on these new draws, plotting the resulting distributions with lower opacity. Fifty iterations are displayed per set of planets. The theoretical projected obliquity distributions are overlaid for the Kozai-Lidov effect with a stellar \citep{naoz2012formation} or planetary \citep{naoz2011hot, petrovich2016warm} companion, secular chaos \citep{teyssandier2019formation}, and planet-planet scattering \citep{nagasawa2011orbital}.}
    \label{fig:lambda_prob_distribution}
\end{figure}

A direct comparison of each examined subpopulation is provided in Figure \ref{fig:lambda_prob_distribution}. The full sample is segmented by temperature (above/below the Kraft break) and by eccentricity, where we consider the circular and eccentric populations separately. {Systems with hot host stars or eccentric planets span a wide range of stellar obliquities.} By contrast, the circular distribution around cool stars, in purple, is heavily weighted towards aligned systems ($|\lambda| \sim 0\degree$).

\section{Tidal Damping in Hot Jupiter Systems}

We considered the effects of tidal damping to investigate the potential origins of the different stellar obliquity distributions in Figure \ref{fig:lambda_prob_distribution}. Once a system becomes misaligned, interactions with the host star continually act to damp that misalignment. All bound planets are affected by interactions with their host stars. However, the extent to which those interactions alter the planet's orbit varies strongly with the system parameters. Stars below the Kraft break have convective envelopes and efficient magnetic dynamos, whereas stars above the Kraft break have radiative envelopes and weaker magnetic braking \citep{dawson2014tidal}. As a result, cool stars can much more efficiently damp out tidal oscillations and realign their companions \citep{winn2010hot}.

In the classical equilibrium tide theory, the tidal realignment timescale for cool stars is given by
\begin{equation}
\tau_{\lambda} = k(M_{\rm cz}/M_p)(a/R_*)^6(1-e^2)^{9/2}(1 + 3e^2 + (3/8)e^4)^{-1},
\label{eq:realign_ecc}
\end{equation} 
where $M_{cz}$ is the mass of the convective envelope, $M_p$ is the planet mass, $R_*$ is the stellar radius, $a$ is the planet semimajor axis, and $k$ is a constant \citep{winn2010hot}. While this model is a simplified heuristic of a more nuanced tidal theory \citep{ogilvie2014tidal, lin2017tidal}, rather than an exact relation, it captures the global properties of the system’s behavior. 

\begin{figure}
    \centering
    \includegraphics[width=1.0\linewidth]{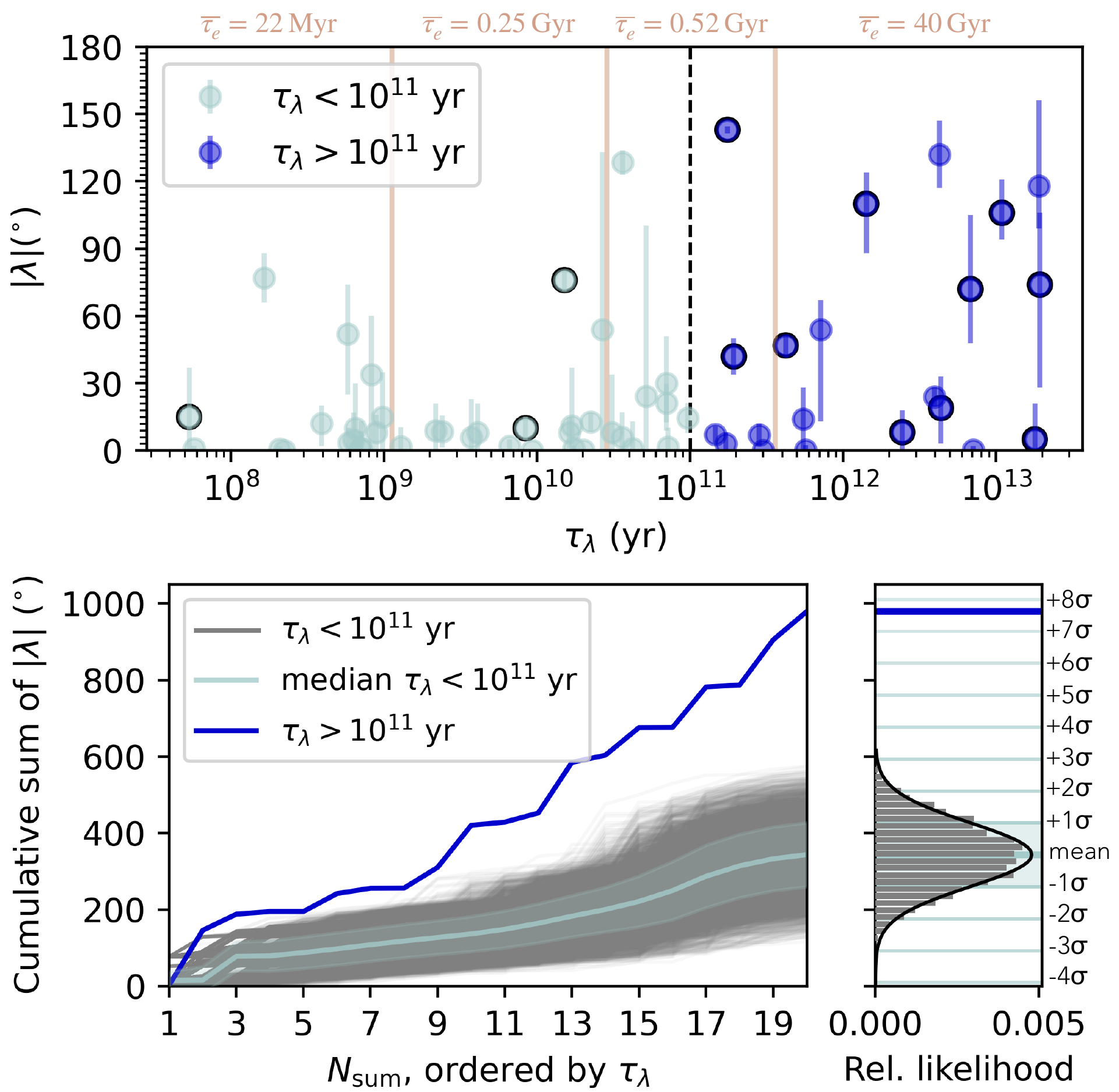}
    \caption{Demonstration that systems with long obliquity damping timescales around cool stars are significantly more misaligned than those with shorter damping timescales. Top: Tidal realignment timescales for cool ($T_{\rm eff} < 6100$ K) star systems in our sample. Planets with $e\geq0.1$ are outlined with a black border. The mean eccentricity damping timescale $\overline{\tau_e}$ for each of the four evenly-sized, 15-planet bins is provided along the top of the panel. Bottom: Stars with the 20 longest tidal realignment timescales ($\tau_{\lambda} > 10^{11}$ years) as compared with random draws without replacement from the population of systems with shorter tidal timescales.}
    \label{fig:winn2010_tides}
\end{figure}

Figure \ref{fig:winn2010_tides} shows the cumulative sum of spin-orbit angles as a function of $\tau_{\lambda}$ for planets orbiting cool stars ($T_{\rm eff} < 6100$ K), with systems segmented by damping timescale (rather than eccentricity as in Figure \ref{fig:cumsum_3panel}). We use $k=10^3$ as a calibrator, such that systems with ages $\tau_{\lambda} < 10^{10}$ years are typically aligned. The measured $|\lambda|$ values were cumulatively summed for the 20 systems with the longest obliquity damping timescales ($\tau_{\lambda} > 10^{11}$ years) and compared to random draws without replacement from the population of planets with shorter timescales $\tau_{\lambda} < 10^{11}$ years. 

As in Figure \ref{fig:cumsum_3panel}, the median of the randomly sampled distribution is shown in light blue together with the region within $1\sigma$ of the median. Systems with $N_{\rm pl} > 2$, which, within our sample, are “peas-in-a-pod" systems \citep{millholland2017kepler, weiss2018california}, were excluded from Figure \ref{fig:winn2010_tides}. To determine the mass of the convective envelope $M_{\rm cz}$, we applied a previously calculated model \citep{pinsonneault2001mass} relating the stellar $T_{\rm eff}$ to $M_{\rm cz}$. All other parameters were drawn from the NASA Exoplanet Archive and supplemented with values from the Extrasolar Planets Encyclopaedia.

The cumulative sum in Figure \ref{fig:winn2010_tides} reveals that, at a $7.6\sigma$ confidence level, planets with longer tidal realignment timescales tend to be observed with larger orbital misalignments. This supports the high-eccentricity migration framework in which hot Jupiter systems around cool stars often begin with large misalignments that are damped over time. Recent work has similarly found that high obliquities of giant exoplanet host stars are almost exclusively associated with wide-separation planets or hot stars \citep{wang2021aligned}, which have long tidal realignment timescales.

High-eccentricity migration is initialized by $N$-body interactions in systems with three or more constituent masses. Dynamical interactions push one planet onto an extremely eccentric orbit, which is gradually recircularized through tidal interactions with the host star. These interactions can simultaneously account for the elevated eccentricities and spin-orbit angles of $e\geq0.1$ exoplanets orbiting stars both above and below the Kraft break. They can also produce orbits with large initial eccentricities and misalignments that have subsequently tidally circularized, and that have, in some cases, realigned. $N$-body mechanisms that are capable of exciting high eccentricities and large spin-orbit misalignments include secular chaos \citep{wu2011secular, hamers2017secular, teyssandier2019formation}, Kozai-Lidov interactions \citep{wu2003planet, petrovich2015steady, anderson2016formation, vick2019chaotic}, and planet-planet scattering \citep{rasio1996dynamical, beauge2012multiple}. 


\begin{figure}
    \centering
    \includegraphics[width=1.0\linewidth]{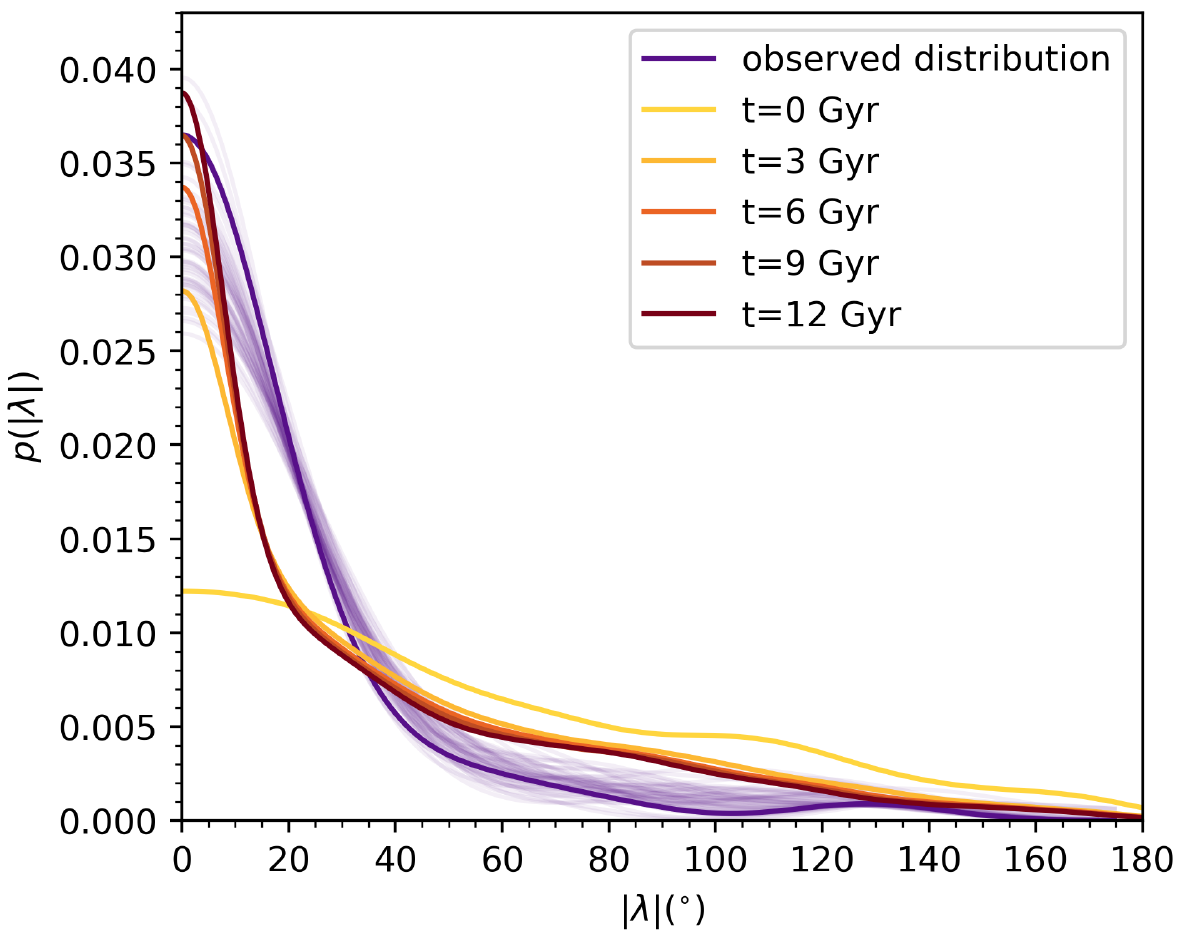}
    \caption{Damping evolution for a sample primordial $|\lambda|$ distribution. The initial obliquity model is comprised of 20\% stellar Kozai-Lidov systems, 40\% planet Kozai-Lidov systems, 10\% secular chaos systems, and 30\% planet-planet scattering systems. Planets are initialized with the same distribution of semimajor axis, planet mass, age, and stellar $T_{\rm eff}$ as the current set of cool stars with $e = 0$ planets and measured $\lambda$.}
    \label{fig:damping_evolution}
\end{figure}

A combination of these processes, together with differential tidal dissipation in hot and cool star systems, can account for the currently observed $| \lambda |$ distributions in Figure \ref{fig:lambda_prob_distribution}. In Figure \ref{fig:lambda_prob_distribution}, theoretical $| \lambda |$ distributions produced by each of these mechanisms are provided alongside the observed distributions. We propose that all four observed distributions in Figure \ref{fig:lambda_prob_distribution} are consistent with an origin from the same set of high-eccentricity formation channels, and that differences in these distributions that are observed today are the natural consequence of obliquity damping. 

To demonstrate the effects of tidal damping, we first focused on the obliquity evolution of $e=0$ planets orbiting cool stars. We applied Equation \ref{eq:realign_ecc} to evolve a mixture model in which 20\%, 40\%, 10\%, and 30\% of planets obtain their obliquities through stellar Kozai-Lidov, planet Kozai-Lidov, secular chaos, and planet-planet scattering, respectively, using the starting distributions in Figure \ref{fig:lambda_prob_distribution}. This distribution is consistent with the high frequency of distant giant perturbers \citep{ngo2015friends, bryan2016statistics} that have been proposed to excite the inclinations of their shorter-period companions \citep{wang2021soles}.

We fit a kernel density estimation (KDE) to the distributions of host star $T_{\rm eff}$, age, $M_p$, and $a/R_*$ for the 33 $e=0$ planets in our sample that orbit stars below the Kraft break, then drew random values from each of these smoothed distributions to produce a set of 10,000 simulated systems. All systems were initialized with $e=0$. We assumed a linear damping rate, and we set $k=10^3$ in in accordance with Figure \ref{fig:winn2010_tides}.

We ultimately found that the distribution evolves along the pathway shown in Figure \ref{fig:damping_evolution} as a result of tidal damping. The theoretical KDE at $t=12$ Gyr shows a peak at low obliquities analogous to that of the observed distribution. Minor discrepancies at moderate $|\lambda|$ values may result from the small number of misaligned planets (five planets with $|\lambda|>35\degree$) that shape the tail of the smoothed, observed sample, without necessarily indicating a true disagreement between the two distributions.

The sky-projected obliquities were directly evolved under the implicit assumption that $\lambda=\psi$. At low $|\lambda|$ values, $|\lambda|$ acts as a lower limit on $|\psi|$, whereas $|\psi|$ is more likely to be close to $|\lambda|$ for larger $|\lambda|$ values \citep{fabrycky2009exoplanetary}. This bias indicates that some fraction of systems observed with low sky-projected obliquities should actually have larger 3D obliquities. Because the probability density peaks at $\lambda=\psi$ even for low measured $\lambda$ values (see Figure 3 of \citet{fabrycky2009exoplanetary}), the distribution should not change dramatically if the true distribution $p(\psi)$ was evolved rather than the sky-projected distribution $p(\lambda)$.

Our proof-of-concept shows that tidal damping can reproduce the current distribution of observed $e = 0$ cool star obliquities based on an initial model comprised of secular mechanisms, without the requirement of invoking disk migration or in-situ formation. We emphasize that we do not rule out contributions from these mechanisms, but, rather, we show that they are not stricly required to account for the stellar obliquity distribution. Minor adjustments to the weighting of secular processes can reproduce the circular hot star distribution and the two eccentric distributions in Figure \ref{fig:lambda_prob_distribution}, each of which has been relatively unaffected by damping. 


\section{The Role of Orbital Eccentricity}

The eccentricities of misaligned systems provide an independent test of high-eccentricity migration. While our analysis up to this point has focused on the stellar obliquity damping timescale $\tau_{\lambda}$, we can also consider the timescale, $\tau_e\sim e/(de/dt)$, for eccentricity evolution driven by tidal dissipation within the planet. Under the effects of tidal dissipation, the evolution of a planet's eccentricity is given by

\begin{equation} 
\frac{de}{dt} = \frac{dE}{dt}\frac{a(1-e^2)}{GMme}\, ,
\end{equation} 
where $M$ is the host star mass, $m$ is the mass of the planet, and $E$ is the orbital energy of the planet. 

The rate of energy dissipation for a synchronously rotating planet is

\begin{equation}
\frac{dE}{dt} = \frac{21k_2 GM^2 \Omega r^5}{2Qa^6}\zeta(e).
\end{equation}
Here, $Q$ is the planet's effective tidal dissipation parameter, $k_2$ is the planet's Love number, and $\Omega$ is the pseudosynchronous rotation rate, given by

\begin{equation}
    \Omega = \frac{1 + \frac{15}{2}e^2 + \frac{45}{8}e^4 + \frac{5}{16}e^6}{(1 + 3e^2 + \frac{3}{8}e^4)(1 - e^2)^{3/2}}n
\end{equation}
where $n$ is the mean motion of the planet's orbit. We set $Q=10^5$ and $k_2=0.3$. The corrective factor $\zeta(e)$, derived in \citet{wisdom2008tidal}, is defined as

\begin{equation} 
\zeta(e) = \frac{2}{7}\Big[\frac{f_0(e)}{\beta^{15}} - \frac{2f_1(e)}{\beta^{12}} + \frac{f_2(e)}{\beta^9}\Big],
\end{equation}

where

\begin{equation} 
f_0(e) = 1 + \frac{31}{2}e^2 + \frac{255}{8}e^4 + \frac{185}{16}e^6 + \frac{25}{64}e^8
\end{equation} 

\begin{equation} 
f_1(e) = 1 + \frac{15}{2}e^2 + \frac{45}{8}e^4 + \frac{5}{16}e^6
\end{equation} 

\begin{equation}
f_2(e) = 1 + 3e^2 + \frac{3}{8}e^4
\end{equation} 

\begin{equation} 
\beta = \sqrt{(1-e^2)}.
\end{equation}

Separating the 60 planets in Figure \ref{fig:winn2010_tides} into four evenly sized bins of 15 planets each, ordered by $\tau_{\lambda}$, we demonstrate that systems with shorter $\tau_{\lambda}$ also preferentially have shorter $\tau_e$ such that their eccentricities and obliquities should be jointly damped over the lifetime of the system. $\tau_e$ is typically shorter than $\tau_{\lambda}$ such that hot Jupiters should often circularize before realigning. In contrast, systems with the longest $\tau_{\lambda}$, which include most of the $e\geq0.1$ population, have $\tau_e$ values exceeding the age of the system. Both timescales, which are contributed by two independent processes -- damping within the planet and damping within the star -- are therefore consistent with the observed distributions. If all systems formed through high-eccentricity migration, the observed $\lambda$ and $e$ distributions would look as they do today.

\section{Implications for Hot Jupiter Formation Theory}

Our results provide two key constraints on the obliquity distribution of Rossiter-McLaughlin targets, which are primarily hot Jupiter host stars. The first constraint, which is a variation on previous findings \citep{winn2010hot, schlaufman2010evidence}, is the observation that stars hosting circular hot Jupiters span a wider range of obliquities above the Kraft break than at lower temperatures. The second is the absence of this pattern in the eccentric sample, where obliquities are consistent with no change at the Kraft break. Together, these observations demonstrate that the population of hot Jupiters is consistent with formation through high-eccentricity migration and suggest that dissipative mechanisms are vital for shaping the obliquity distribution of hot Jupiter host stars.

The absence of a significant change in obliquities at the Kraft break for eccentric systems indicates that dissipative mechanisms have not had time to sculpt the eccentric population in the same way that they have shaped the circular population. In the framework of Kozai capture \citep{naoz2012formation}, eccentric planets are either experiencing ongoing eccentric Kozai-Lidov oscillations or they have had these oscillations suppressed by apsidal precession in the past. Our sample includes well-characterized systems such as that of HD 80606, in which the transiting planet's high obliquity and eccentricity are cleanly recovered through Kozai migration \citep{wu2003planet}.

A subset of eccentric planets that have two or more planetary companions may be undergoing secular chaos. Secular chaos transfers angular momentum outwards due to the overlap of resonances in a multiplanet system, elevating the orbital inclination and eccentricity of the innermost planet. In either excitation framework, tidal dissipation, which acts differentially in hot and cool star systems, has not had time to globally alter the $|\lambda|$ distribution of the eccentric population.

For hot Jupiter systems with circular orbits, tidal dissipation has played a more important role in shaping the currently observed stellar obliquity distribution. These systems are consistent with $N$-body interactions that were suppressed early on such that the companion orbits were able to fully tidally circularize and, in some cases, realign. Planets orbiting cool stars realign quickly, while those orbiting hot stars have much longer tidal realignment timescales and remain closer to their primordial spin-orbit angles.

Systems with both large misalignments and high eccentricities may instead be produced by a combination of primordial disk misalignments and planet-planet or planet-disk interactions \citep{duffell2015eccentric, anderson2017moderately, frelikh2019signatures, anderson2020situ, debras2021revisiting}. Our results do not rule out these alternative scenarios, but, rather, they provide a relatively simple framework that is fully consistent with the observed $\lambda$ and $e$ distributions. Previous work has revealed that long-period ($P>5$ day) planets orbiting hot stars tentatively demonstrate a trend towards alignment \citep{rice2021soles}. Because these planets are exceptionally difficult to realign through tidal interactions, this trend, if confirmed, may suggest that protoplanetary disks are typically aligned and that the misalignments of hot Jupiters are attained through dynamical interactions after the disk has dispersed.



\section{Conclusions}
\label{section:conclusions}
Our analysis establishes that the observed distribution of hot Jupiter host star obliquities can arise naturally from a combination of high-eccentricity migration and obliquity damping mechanisms. Cool stars hosting circular planets have had the most strongly damped obliquities, while hot stars and hosts of eccentric planets have experienced weaker damping. We predict that, under our proposed framework, the observed difference between the eccentric and $e = 0$ cumulative sums will grow with additional observations. We conclude that the stellar obliquity distribution for hot Jupiter systems is consistent with having been crafted primarily by high-eccentricity migration and tidal damping, with no requirement to appeal to disk migration or in-situ formation at the population level.

\section{Acknowledgements}
\label{section:acknowledgements}

M.R. is supported by the National Science Foundation Graduate Research Fellowship Program under Grant Number DGE-1752134. This research has made use of the NASA Exoplanet Archive, which is operated by the California Institute of Technology, under contract with the National Aeronautics and Space Administration under the Exoplanet Exploration Program.

\software{\texttt{numpy} \citep{oliphant2006guide, walt2011numpy, harris2020array}, \texttt{matplotlib} \citep{hunter2007matplotlib}, \texttt{pandas} \citep{mckinney2010data}, \texttt{scipy} \citep{virtanen2020scipy}, \texttt{emcee} \citep{foremanmackey2013}}

\facility{Exoplanet Archive, Extrasolar Planets Encyclopaedia, Open Exoplanet Catalogue}

\appendix 

\section{Adopted Parameters}
\label{adopted_parameters}

Our full samples of parameters, drawn from archival studies, are provided as supplementary data for Figure \ref{fig:cumsum_3panel}. All stellar and planetary parameters other than stellar multiplicity and $\lambda$ were drawn directly from the NASA Exoplanet Archive, with ages supplemented by the Extrasolar Planets Encylopaedia. The stellar multiplicity of each system, provided for reference, was determined through cross-matching with the Catalogue of Exoplanets in Binary Star Systems \citep{schwarz2016new} and the Open Exoplanet Catalogue.

\section{Orbital Period vs. Eccentricity}
One alternative possibility is that eccentricity acts as a proxy for a different trend in the dataset. The tidal circularization timescale of a short-period planet scales with semimajor axis as $\tau_{\rm cir} \propto a^{13/2}$ \citep{murray1999solar}, meaning that small differences in semimajor axis correspond to dramatically different tidal circularization timescales. Planets on eccentric orbits, by definition, have not completed the tidal circularization process. This means that they may also tend to have larger semimajor axes, or, equivalently, longer orbital periods ($P$) as compared with $e=0$ planets. 

\begin{figure}
    \centering
    \includegraphics[width=1.0\linewidth]{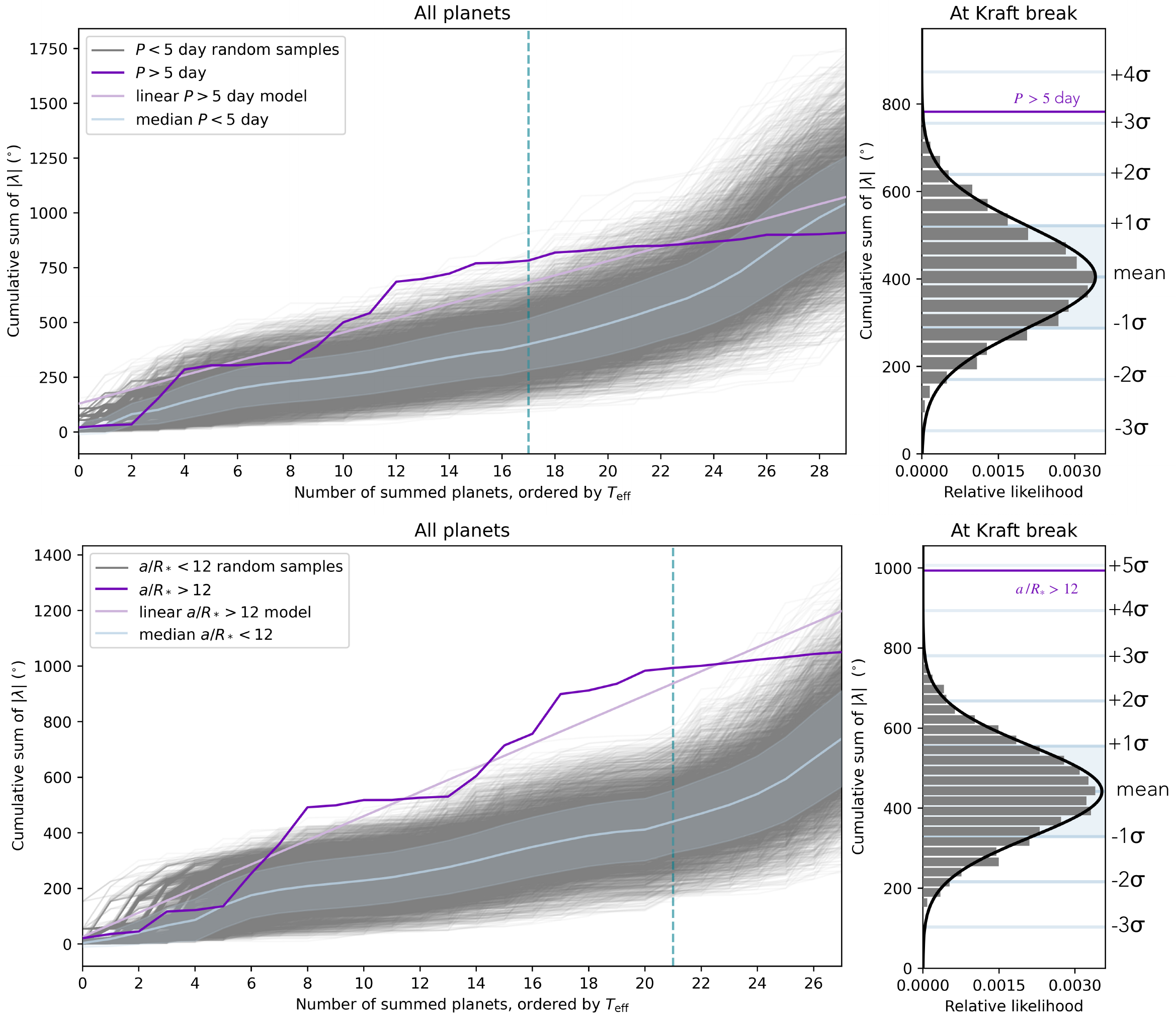}
    \caption{Cumulative sums as a function of orbital period $P$ (top) and orbital separation $a/R_*$ (bottom), for comparison with the eccentricity cumulative sum in Figure \ref{fig:cumsum_3panel}. }
    \label{fig:cumsum_byperiod_and_aoverRstar}
\end{figure}

To address this possibility, we carried out the same analysis as a function of orbital period (comparing the $P>5$ day and $P<5$ day populations) and as a function of orbital separation (comparing the $a/R_*>12$ and $a/R_*<12$ populations), with results shown in Figure \ref{fig:cumsum_byperiod_and_aoverRstar}. If the observed effect is predominantly due to a correlation between obliquity and orbital period (orbital separation), rather than eccentricity, the population should show a stronger increase in misalignments with increasing $P$ ($a/R_*$) than $e$.

In both cases, the significance of our result was substantially weaker than when dividing the sample by eccentricity ($6.5\sigma$), with only a $3.3\sigma$ signal when segmenting by $P$ and a $4.9\sigma$ signal when segmenting by $a/R_*$. Figure \ref{fig:cumsum_byperiod_and_aoverRstar} also shows substantial structure below the Kraft break in the $P>5$ day and $a/R_* > 12$ sums. This suggests that divisions by $P$ or $a/R_*$ may produce a more heterogeneous population than that produced by our eccentricity cut, where planets smoothly follow a relatively consistent upward trend in $\sum|\lambda|$. 

Finally, both panels of Figure \ref{fig:cumsum_byperiod_and_aoverRstar} reveal a nearly flat trend in $\sum|\lambda|$ above the Kraft break. This demonstrates that relatively long-period systems around hot stars have typically been observed to be aligned, as pointed out in \citet{rice2021soles}.

\bibliography{bibliography}
\bibliographystyle{aasjournal}

\end{document}